# A new sample of bright galaxy pairs in UZC[⋆]

P.Focardi[1], V.Zitelli[2], S.Marinoni[2], B.Kelm[1]

[1] Dipartimento di Astronomia, Universitá di Bologna, Italy
    e-mail: paola.focardi@unibo.it
[2] INAF-OABO, Via Ranzani 1, 40127 Bologna, Italy

**ABSTRACT**

*Aims.* We present a new sample of bright galaxy pairs extracted applying an objective selection code to UZC catalog. The sample is volume limited to $M_{sw}$ = -18.9 +5 log $h$ and contains 89 galaxy pairs.
*Methods.* We analyze the kinematical, morphological and photometrical properties of galaxies belonging to this sample.
*Results.* We show that velocity separation, $|\Delta v_r|$, between pair members is significantly lower in spiral type (S+S) pairs than in early-type (E+E) and mixed (E+S) pairs. This indicates that truly isolated galaxy pairs are more likely to be found among S+S pairs. We show that ellipticals are rare and underluminous in B and that late spirals (T ≥ 4) are overluminous.
*Conclusions.* We confirm that the formation of bright ellipticals is a phenomenon linked to group/cluster environment, while galaxy-galaxy interaction may enhance blue luminosity of disk galaxies through SF phenomena. This last statement is supported by the presence of strong FIR emission from early spirals in this sample and by the high frequency of AGN/SB phenomenon, revealed mainly in pairs of low relative radial velocity separation and showing signs of interaction.

**Key words.** galaxies: general – galaxies: fundamental parameters – galaxies: interaction

## 1. Introduction

It is well known and accepted that starburst (**SB**) activity can be triggered, and possibly enhanced, by galaxy encounters. On theoretical basis (Barnes & Hernquist 1991) galaxy interaction is expected to redistribute large amount of material towards the galaxy central regions, which may consequently trigger and fuel violent burst of star formation. Galaxy collisions, eventually leading to final merging, can produce strong alterations of galaxy morphologies and even originate complex structures, such as tidal tails and bridges. Instabilities in the discs, arising from tidal interactions, may induce formation of galaxy bars (Noguchi 1988; Barnes & Hernquist 1991) which generating an inflow of gas towards the galaxy central regions might even activate AGN phenomenon. This theoretical scenario, however, is not adequately supported by observations. Eventhough, starting from Larson & Tinsley (1978) there has been growing evidence, obtained on several different samples, of an increase of star formation in interacting galaxy systems (e.g. Kennicut & Keel 1984; Kennicut 1987; Keel 1993, 1996; Donzelli & Pastoriza 1997; Barton et al. 2000) a one-to-one correlation between galaxy-galaxy interaction and star formation remains not obvious. In fact only an extremely limited number of objects (ULIRGs, Sanders & Mirabel 1996) show a fraction of interacting galaxies nearly close to 100% (Sanders et al 1988; Borne et al. 1999) while there are several interacting systems showing no sign of star formation. The situation becomes even more complex and controversial for the so called AGN-interaction paradigm for which conflicting results have been given so far (Dahari et. al. 1985; Keel et al. 1985; Fuentes-Williams & Stocke 1988; Rafanelli et al. 1995; Macckenty 1989; Kelm et al. 1998; De Robertis et al 1998; Schmitt 2001; Kelm et al 2004). However these results relate to samples which often are small, have been selected by different methods and criteria and may be biased towards or against certain kind of systems. Large part of controversy is thus likely to be ascribed to selection inhomogeneities among the samples.

Galaxy pairs are the ideal sites in which to investigate the role of a close companion on galaxy formation and evolution. So far, two large samples of "nearby" galaxy pairs are available: the Karactchensev sample ( KPG, Karachentsev 1972) and its southern counterpart (RR, Reduzzi & Rampazzo 1995). The first one has been selected by visual inspection of POSS plates, the second one has been extracted applying KPG criteria to the ESO-LV catalogue (Lauberts & Valentijn, 1989).

Recent availability of large complete nearby 3D galaxy catalogues has made possible to select galaxy samples having well defined environment characteristics (Focardi & Kelm, 2002). We present here a volume limited new sample of 89 bright galaxy pairs (UZC-BGP) selected applying an objective algorithm to UZC catalog (Falco et al. 1999).This sample does not suffer of velocity/distance biases and of contamination by projection effects as galaxies are already close in 3D space. In this paper we present the UZC-BGP sample (§2) and, on the basis of presently available data, discuss its morphological content and show how it relates to the radial velocity separation ($\mid \Delta v_r \mid$) between galaxies in the pairs (§3). We analyze B luminosity and NIR colors (§4), FIR emission (§5) and the presence of nuclear activity (AGN/SB) (§6). The conclusions are drawn in §7. A Hubble constant of $H_0 = 100\,h$ km s$^{-1}$ Mpc$^{-1}$ is assumed throughout.

## 2. The sample

The bright pair galaxy sample has been selected with an adapted version of the neighbour search algorithm of Focardi & Kelm (2002). Each UZC galaxy having $M_{zw} \le -18.9 + 5$ log $h$ , $v_r \in$ [2500-7500] km s$^{-1}$ and $\mid b^{II} \mid \ge 30^o$ has been explored on a surrounding area characterized by a projected radius $R_p$ = 1 $h^{-1}$ Mpc and a radial velocity "distance" $\mid \Delta v_r \mid \le 1000$ km s$^{-1}$. The limit in $\mid b^{II} \mid$ has been set to minimize the effects of galactic absorption. The lower limit in $v_r$ reduces distance uncertainty due to peculiar motions and avoids contamination by the Virgo cluster. The upper limit in $v_r$ results by the combination of UZC magnitude limit ($m_{zw} \le 15.5$) and our imposed limit in $M_{zw}$. The radial velocity "radius" ($\mid \Delta v_r \mid$) has been set equal to 1000 km s$^{-1}$, which is large enough to not induce an artificial cut in relative velocity of galaxies in pairs and to prevent contamination by galaxy groups. Galaxies having only one luminous ($M_{zw} \le -18.9 + 5$ log $h$ ) neighbour within $r_p \le 200\,h^{-1}$ kpc and no other ones up to $R_p \le 1\,h^{-1}$ Mpc are part of the bright galaxy pair sample (UZC-BGP). The galaxy-galaxy projected distance radius ($r_p$) has been set equal to 200 $h^{-1}$ kpc to account for possible huge haloes tied to bright galaxies (Bahcall et al. 1995; Zaritsky et al. 1997). The large scale radius ($R_p$) has been set equal to 1 $h^{-1}$ Mpc to ensure absence of luminous galaxies on the scale typical of groups.

Application of the above described selection criteria has produced a sample of 89 galaxy pairs, listed in Table 1 (available at the CDS). For each galaxy we give pair and galaxy identification (column 1 and 2), equatorial coordinates RA$_{J2000}$ and DEC$_{J2000}$ (column 3 and 4), apparent magnitude m$_{zw}$ and radial velocity $v_r$, from UZC, (columns 5 and 6), morphological classification from LEDA (column 7). For 18 galaxies morphological classification is not too accurate thus it is indicated by an E or a S followed by a question mark. The presence of a ring is denoted by Y (yes) in column 8 and the presence of nuclear activity , from available literature data (NED), is indicated in column 9. Column 10 indicates if the galaxy has already been identified as a member of either a galaxy pair or a compact group.

In eleven cases and as consequence of the imposed limit in absolute magnitude the algorithm has detected only the most luminous galaxies of a compact group (HCG, Hickson 1982, Hickson et al. 1992; KTG, Karachentseva et al. 1979; UZC-CG, Focardi & Kelm 2002). For the same reason only the brightest galaxy of 6 known pairs (KPG) have been detected. In these last cases, however, the algorithm has detected a bright companion which KPG visually based criteria failed to identify. All previously identified galaxy pairs (19) fulfilling our selection requirements have been detected by the algorithm. The remaining 53 pairs are new (not identified before).

Having applied well defined selection criteria to a large and complete galaxy catalogue we can compute the fraction of galaxies in pairs which turns out to be 6 %. Luminous galaxies are rare, as the vast majority (88 %) of UZC-BGP has M$_{zw}$− 5 log $h$ > -20. Thus galaxy pairs do not appear ideal sites in which bright galaxies can be originated, contrary to what seems to be the case for galaxy groups especially for elliptical dominated ones (Kelm & Focardi 2004). The magnitude difference ($\mid \Delta M_{zw} \mid$) between pair members is less than 0.5 for the majority (76 %) of galaxies. Most pairs (72 %) display a $\mid \Delta v_r \mid \le 200$ km s$^{-1}$ , while more than half pairs (64 %) have r$_p$ < 100 $h^{-1}$ kpc.

## 3. Morphological content of the sample

Morphological classification is available (LEDA) for 164 galaxies (92% of the total sample); 79 pairs possess classification for both members, 6 for only one and 4 lack this information for both members. As LEDA provides also the morphological code (T) we have used this one to classify these 79 UZC-BGP pairs with morphological classification. Of these 10 are E+E pairs (composed of early-type galaxies only, both with T < 0), 38 are S+S (composed of spirals only, both with T $\ge$ 0) and 31 are E+S pairs (one galaxy with T < 0 and the other with T $\ge$ 0). This gives a morphological content of our sample of 13 % E+E, 48 % S+S and 39 % E+S, to be compared to KPG corresponding figures which are 14 %, 60 % and 26 % respectively. While early-type pair content of UZC-BGP matches extremely well the KPG one, the fraction of mixed and spiral type pairs are rather different, UZC-BGP having a higher content of E+S (and consequently a lower content of S+S) than KPG sample. We stress however that 8 E+S UZC-BGP contain either a SB0 or a S0/a galaxy which (according to its T value just below 0) has been counted as an early-type galaxy. Actually these galaxies are border line objects and including them among early-type or spirals is somewhat arbitrary. Had we adopted a less conservative criterion and included those pairs in the S+S ones would have changed the fraction of UZC-BGP S+S and E+S pairs into 58 % and 29 %, in much better agreement with KPG morphological content. No significant difference emerges either in magnitude or in projected distance when morphological composition of the pairs is taken into account. The radial velocity separation appears instead significantly different (Fig. 1). S+S pairs display a distribution narrower than E+S and E+E ones,while, curiously unclassified pairs distribute similarly to S+S. A Kolmogorov Smirnov test confirms the difference (99.3 % c.l.) in $\mid \Delta v_r \mid$ distribution of E+E and S+S pairs and of S+S and E+S pairs (99.5 % c.l.). Also the $\mid \Delta v_r \mid$ distribution of unclassified pairs is significant different from the one of E+E and E+S (99.2 % c.l. in both cases) The wider radial velocity separation of E+E and E+S might indicate the presence of a deeper potential well, implying that a significant fraction of these pairs could be part of larger bound structures. If this were the case, truly isolated galaxy pairs would be more likely found among S+S pairs only.

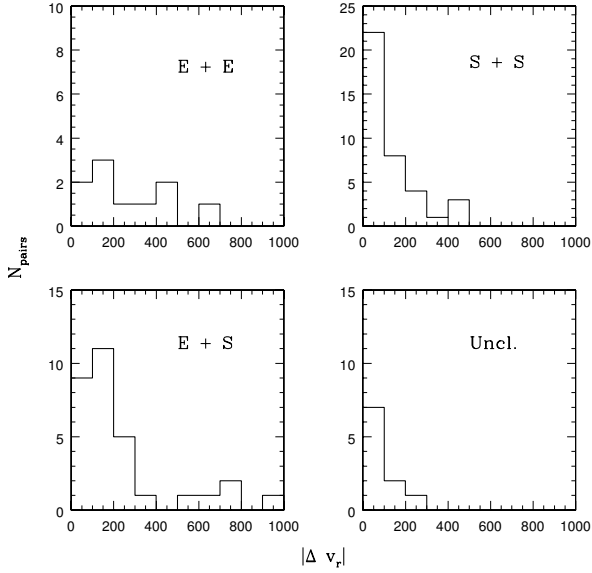

**Fig. 1.** The radial velocity separation, for E+E (upper left), S+S (upper right), E+S (lower left) and unclassified (lower right) pairs.

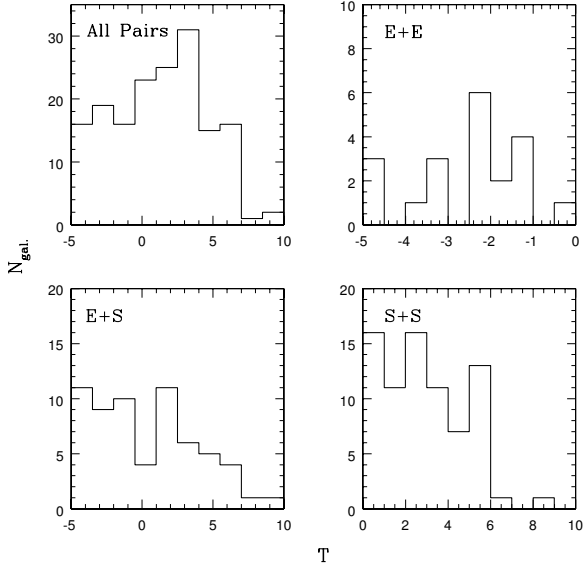

**Fig. 2.** Morphological content of the 79 UZC-BGP pairs having classification for both members (upper left).The other panels show the content of the 10 E+E ,31 E+S and 38 S+S pairs.

Figure 2 shows the morphological type code (T) distribution of galaxies in the 79 UZC-BGP pairs having classification for both members (upper left panel). The other panels show morphological content of E+E (upper right), E+S (lower left) and S+S (lower right). Figure 2 (upper left) illustrates that UZC-BGP sample is dominated by early spirals ($0 \leq T < 4$). Late spirals ($T \geq 4$) are much more rare while early-type galaxies ($T < 0$) are quite abundant. The upper right panel of Fig. 3 shows also that ellipticals ($T < -3$) in early-type pairs are not as frequent as S0s ($-3 \leq T < 0$) but the low statistic does not allow to draw definitive conclusion on this point.

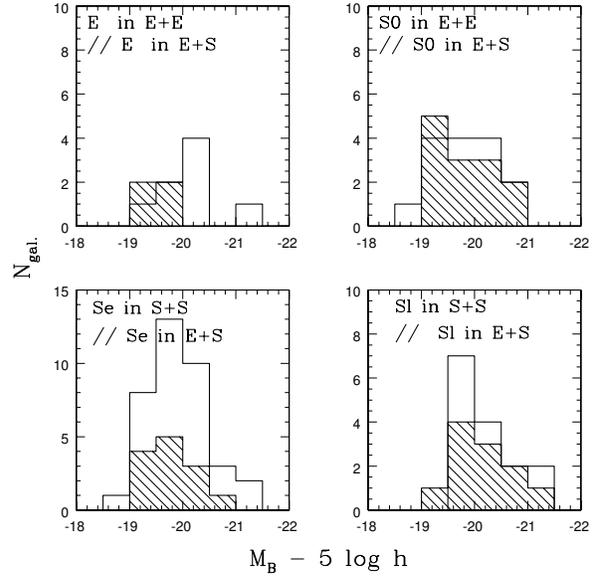

**Fig. 3.** $M_B$ distribution of E (upper left) S0s (upper right), early (lower left) and late (lower right) spirals. The dashed histograms represent the $M_B$ distribution of galaxies belonging to minority population, i.e. E+E in the upper panels, E+S in the lower ones.

## 4. Luminosity and colors of BGP member galaxies

We have collected (from NED) optical and NIR photometric data for UZC-BGP galaxies. B magnitude is available for 128 galaxies and J H K (2-Mass) magnitudes for 168 galaxies ( respectively 72 % and 94 % of the total sample). V magnitude is available only for 21 % of the sample and will not be used.

Figure 3 shows $M_B$ distribution of E and S0s (upper panels) and early and late spirals (lower panels) in pairs of different morphology. The shaded distribution refers always to minority populations which are E+E pairs for early-type galaxies (upper panels) and E+S for spirals (lower panels). The solid distribution relates to E+S pairs (upper panels) and to S+S pairs (lower panels). The upper left panel of Fig. 3 shows that ellipticals are underluminous in B a tendency which appears especially strong when these galaxies are members of E+E pairs. On the other hand spiral galaxies show a luminosity distribution which extends at higher values. This is particularly true for late spirals (lower right panel). Galaxy pairs appear thus characterized by a population of "underluminous" ellipticals and of "overbright" late spirals, implying that this kind of environment might disfavour the formation of bright early-type galaxies and favour the luminosity increase of late spirals. The latter phenomenon beeing probably linked to a SF enhancement induced by interaction processes. No significant differences are found among spiral galaxies in different kind of pairs eventhough the lower left panel of Fig. 3 indicates that the early spirals of higher luminosity are found in S+S pairs. This might indicate an enhancement in luminosity due to interaction phenomena occurring mainly in S+S pairs.

Figure 4 shows the color magnitude diagram (B-K vs. $M_K$) of E and S0s (upper panels) and early and late spirals (lower panels). Symbols are filled when galaxy morphology accords with pair one , void when it does not (i.e. open symbols represent always E+S pairs, filled circles in the upper panels represent E+E pairs, filled symbols in the lower ones represent S+S pairs ). Spirals (lower panels) distribute in the usual way, the brightest in K being the reddest. Late spirals are significantly bluer than early spirals. This happens as a consequence of the B luminosity distribution of late spirals which is shifted of about one magnitude with respect to the one of early spirals (Fig. 4). Early-type galaxies are bluer than normal (< B-K >$_{E-S0}$ ~ 4.5). S0s (upper right panel) appear particularly concentrate while the sequence of ellipticals (upper left panel) is much extended. The K brightest ellipticals in E+S pairs (open circles) deviate from the normal sequence being bluer than expected. This happens as a consequence of their being also the brightest in $M_B$ (see Fig.3). We thus find that ellipticals in pairs display anomalous characteristics which appear to relate to pair morphology: ellipticals in E+E are underluminous in B (see Fig. 3), while ellipticals of higher B luminosity, in E+S pairs, are bluer than expected (Fig.4). This finding suggests that interaction phenomena (eventhough limited) affecting ellipticals in pairs are more likely to occur in E+S than in E+E which appear to host normal few luminosity ellipticals.

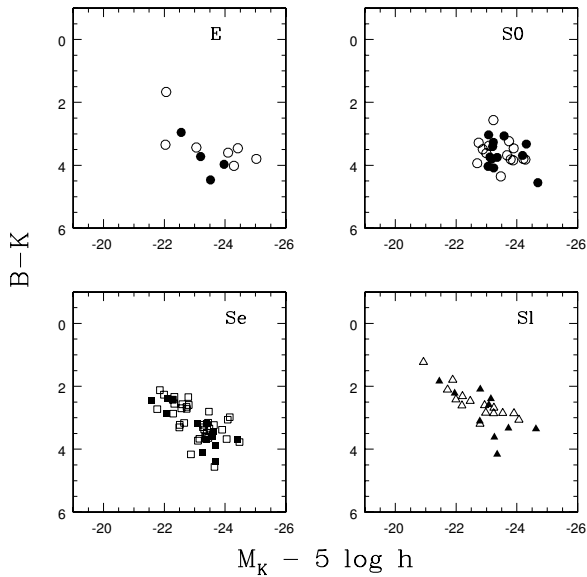

**Fig. 4.** B-K versus $M_K$ distribution. The upper panels show the distribution of E and S0 in E+E (filled circles) and E+S pairs (empty circles). The lower panels the distribution of early and late spirals , filled symbols indicate spirals in S+S while open symbols spirals in E+S pairs.

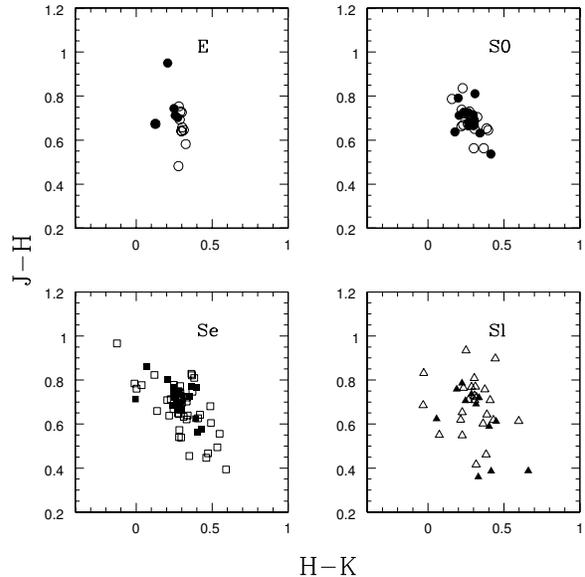

**Fig. 5.** J-H versus H-K distribution. The upper panels show the distribution of E and S0 in E+E (filled circles) and E+S pairs (empty circles). The lower panels the distribution of early and late spirals, filled symbols indicate spirals in S+S, open symbols spirals in E+S pairs.

Figure 5 shows the NIR color-color (J-H vs. H-K) diagram, for E, S0 early and late spirals in pairs whose morphology is either accordant (filled symbols) or discordant (open symbols) with the galaxy one. The upper left panel confirms that ellipticals in pairs have bluer than normal colors. In fact all but one (belonging to an E+E pair) are bluer than J-H = 0.78, which is the normal value for 'ordinary' (bulge dominated) galaxies (Glass 1984) and all but two (always belonging to E+E pairs) have H-K larger than 0.22 which is again (Glass 1984) the normal value for bulge dominated galaxies. The distribution of spirals (lower panels) is significantly shifted towards bluer J-H and redder H-K colors. Most galaxies have H-K > 0.26 which is typical of emission line galaxies (Glass 1984; Glass & Morwood 1985) and the distribution of late spirals (lower right panel) is much scattered (in H-K) than the distribution of early spirals (lower left panel).

## 5. Far Infrared emission from UZC-BGP

We have collected (from NED) IRAS data for all galaxies in our sample. As expected, the fraction of FIR emitting galaxies increases as the morphological type advances. Table 2 (column 2) shows the morphological content, in ellipticals, S0s, early and late spirals, of the UZC-BGP sample, and the corresponding number of galaxies having FIR emission measured in all 4 IRAS bands (column 3). In column 4 is indicated the number of galaxies having measured $f_{60}$ and $f_{100}$ and an upper limit estimate for $f_{12}$ and/or $f_{25}$. In column 5 the number of galaxies having only upper limit estimates either for all the 4 bands or for at least one of the shortest ($f_{12}$ ,$f_{25}$) and of the longest ($f_{60}$, $f_{100}$) wavelength bands.

The fraction of IRAS detected galaxies turns out to be 22% (4/18) for ellipticals, 31 % (11/35) for S0s, 46 % (33/72) for early spirals and 59 % (23/39) for late spirals which is rather high (especially for early-type galaxies). These fractions decrease to 5 %, 20 %, 38 % and 49 % considering only galaxies for which is possible to compute $L_{FIR}$ (column 3 and 4 of Table 2). However, the computed fractions must be considered lower limit estimates of the real ones as faint galaxies easily go undetected in a flux limited survey as IRAS. If we keep UZC-BGP galaxies up to $m_{zw} \leq 14.5$ and recompute the fractions we find that galaxies detected and having $L_{FIR}$ measurable are 40 % and 10 % for E, 47 % and 27 % for S0s, 62 % and 53 % for early spirals and 79% and 64% for late spirals.

**Table 2.** IRAS emission from UZC-BGP galaxies

| Morphological Code | $N_{Tot}$ | $N_{IRAS}$ | $N_{60-100}$ | $N_{ul}$ |
|---|---|---|---|---|
| T < -3 | 18 | 0 | 1 | 3 |
| -3 ≤ T < 0 | 35 | 1 | 6 | 4 |
| 0 ≤ T < 4 | 72 | 14 | 13 | 6 |
| T ≥ 4 | 39 | 6 | 13 | 4 |

Figure 6 shows the IRAS color color diagram for ellipticals (upper left panel), S0s (upper right), early (lower left) and late (lower right) spirals. Circles represent galaxies for which FIR emission has been measured in all 4 IRAS bands. In analogy with Fig. 5 and 6 they are filled if the galaxy resides in an accordant morphology pair (E+E for the upper panels, S+S for the lower ones) and they are void if the galaxy resides in an E+S pair. Triangles represent galaxies for which both $f_{60}$ and $f_{100}$ have been measured (while an upper limit estimate of either both or at least one of $f_{12}$ and $f_{25}$ is available). They may be either filled or void according to the above stated definition. The high number of filled circles in the lower left panel confirms that galaxy-galaxy interaction, occurring mainly in S+S pairs enhance FIR emission in early spirals. More than half early spirals lay above $\log( f_{60}/f_{100}) > -0.3$ suggesting the presence of SF phenomena (Sanders & Mirabel 1996; Kim & Sanders 1998). The two early spirals displaying the highest value of $f_{60}/f_{100}$ are NGC 7771 which is a known SB and NGC 5908. They belong to pairs having both member emitting in FIR. Late spirals show a much concentrated distribution and do not appear to 'prefer' S+S pairs to E+S ones. In Fig. 6 we have not displayed galaxies which have an upper limit estimate of the flux in at least one of both the short and the long IRAS wavelenght band. If we take into account also these galaxies we find that FIR emission is detected in only 3 early-type galaxies, belonging to 3 distinct E+E pairs, in 22 E+S pairs (5 of which showing emission for both members) and in 27 S+S pairs (9 of which showing emission for both members). This gives a frequency of FIR emission of 30% for E+E and 69% for both E+S and S+S. There is no significative difference in the distribution of $|\Delta v_r|$ and $r_p$ between UZC-BGP hosting and non hosting one (or two) IRAS sources eventhough the former show a narrower $|\Delta v_r|$ distribution than the latter.

We have computed Far Infrared Luminosity ($L_{FIR}$) (following Sanders & Mirabel 1996) for galaxies having $f_{60}$ and $f_{100}$ measured and compared it to $L_B$. Most galaxies follow the $\log(L_{FIR})$ - Log ($L_B$) relation , only 3 early-type galaxies deviate significantly from it, of these the one showing the lowest FIR and B luminosity is the only FIR emitting elliptical. The two strongest FIR sources are the brightest in B early spirals (Fig. 4 lower left panel). No LIRGs are present.

## 6. Nuclear activity in UZC-BGP

We have inspected available data (NED) looking for the presence of nuclear activity in UZC-BGP. This information is provided in column 9 of Table 1. The sample contains 10 SB galax-ies, 4 Liners, 3 Sy2, 1 Sy1 and 2 AGN. (In cases of double classification, e.g. Sy2/SB, we have retained the first one). Adding together all kind of activities we find that the incidence of nuclear activity phenomenon in UZC-BGP galaxies is 11% which is quite high considering that it must be taken as a lower limit estimate. SB activity appears to be the most frequent one.

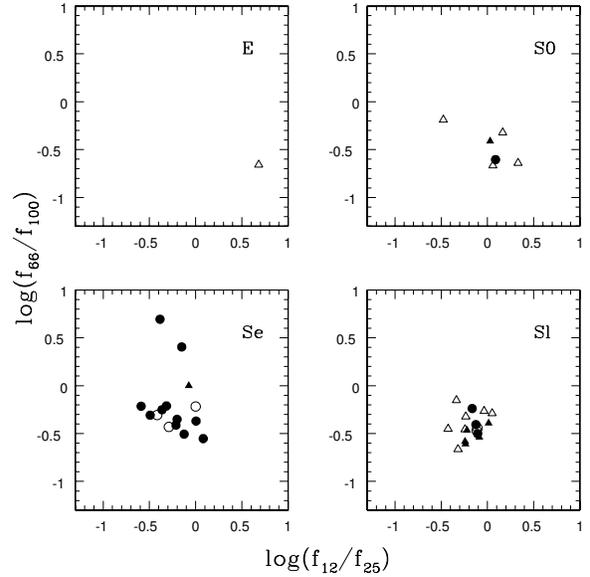

**Fig. 6.** IRAS color color diagram, for ellipticals (upper left), S0s (upper right), early (lower left) and late (lower right) spirals. Symbols are either filled or void in analogy with definition adopted for Fig. 4 and Fig.5 Circles stand for galaxies having FIR emission measured in all 4 IRAS bands, triangles for galaxies having only $f_{60}$ and $f_{100}$ measured and an upper limit estimate for either both, or at least one, of $f_{12}$ and $f_{25}$.

Four pairs have both members active: BGP 1, BGP 85, BGP 87 and BGP 89 Three of these pairs are E+S and one (BGP 89) is a S+S. The pairs hosting only one active members are 3 E+S, 8 S+S and 1 unclassified. In all, but three cases (BGP 73A, BGP 80B and BGP 82A), active galaxies are first ranked (or equal luminosity) in the pair. The vast majority of active galaxies are spirals which is not surprising due to the large incidence of SB activity in this sample. Only 3 are early-type, namely NGC 6251 (BGP 79A) which is an elliptical, NGC 7550 (BGP 85A, E/S0) and NGC 7679 (BGP 87A, SB0/a). All of them are members of E+S pairs. Half of the active galaxies have $M_{zw} \leq -19.5$ -5 log $h$ and are thus among the brightest in the sample. The most luminous active galaxy is the SB NGC 23 (BGP 2A) with $M_{zw}$ - 5 log $h$ = -20.8. The three Sy 2 galaxies have luminosities normal/high ($-20.3 \leq M_{zw}$ - 5 log $h \leq -19.3$), the only Sy 1 detected has $M_{zw}$ - 5 log $h$ = -20.4. The radial velocity separation $|\Delta v_r |$ of 14/16 pairs with one (or both) active members is less than 200 km s$^{-1}$ . The two pairs exceding this limit are BGP 79 (hosting a Sy 2) and BGP 85 (hosting two AGNs). If nuclear activity is related to galaxy-galaxy interaction, pairs having $|\Delta v_r| \leq 200$ km/s appear as the ones in which this phenomenon is most likely to occurr.

We have inspected the DSS images of UZC-BGP hosting one (or both) active member(s) to look for signs of morpholog-ical distorsion and interaction. Strong interaction occurs in all the 4 UZC-BGP pairs with both members active, namely the two Liners in BGP 1 which are part of a well known system of interacting galaxies (VV 254, the Taffy system), the two AGNs in BGP 85 (also part of HCG93), the Sy1/SB and Sy2 in BGP 87 (VV 329) and the SB/Liner and SB in BGP 89. Interaction between pair members is also present in BGP 7 and BGP 14 both hosting a SB galaxy. BGP 28A (Liner) and 38A (SB) are mergers, the last one in a less advanced stage. BGP 73A (SB/Sy1), 80B (SB) and 83A (Sy2/SB) show signs of morphological distorsion. While the most luminous active galaxy of the sample (NGC 23 SB in BGP 2) has a large bar but not visible interaction, the Liner in BGP 3, the SB in BGP 68, the Sy2/Radiogalaxy in BGP 79 and the SB in BGP 82 show no signs of distorsion. Thus 15/20 active galaxies in UZC-BGP show signs of interaction/distorsion providing support to the interaction activity scenario.

## 7. Conclusions

We have presented a new volume limited sample of bright galaxy pairs selected from UZC catalog by means of an objective neighbour-search algorithm. The sample contains 89 pairs whose members have a mutual projected distance $r_p \leq 200\ h^{-1}$ and no other bright ($M_{zw} \leq -18.9 + 5 \log h$) galaxies within $R_p \leq 1\ h^{-1}$ Mpc and $\mid \Delta\ v_r \mid \leq \mathbf{1000}$ km s$^{-1}$.

We have shown that elliptical galaxies are rare (10 %) in pairs and underluminous in B. The last characteristic holds especially for E in E+E pairs ($M_B < -20\ +5\ \log h$) Confirming previous findings (Kelm & Focardi, 2004) relating the formation of bright ellipticals to phenomenon linked to the formation of group size systems. Analysis of NIR data provide evidence that ellipticals in E+S pair are 'bluer' (in J-H and in B-K) than expected. Thus, for elliptical galaxies the effect of a close bright companion appears to relate to its morphology having the maximum (even if limited) effect if it is a spiral. Curiously in the B-K vs, M$_K$ plot S0 galaxies appear extremely concentrated suggesting that S0s in pairs might constitute a population of galaxies with well defined and unique characteristics.

Spiral galaxies, particularly late spirals, have B luminosity enhanced and we suggest that this may be due to galaxy-galaxy interaction phenomena, occurring mainly in S+S pairs. These pairs display the narrowest radial velocity separation between members a characteristic which make them to appear the best candidates of truly isolated galaxy pairs.

Analyzing FIR (IRAS) emission we have shown that it is particularly strong for early spirals, which are the galaxies of highest FIR luminosity eventhough none of them reaches the LIRG level. More than half of the early spirals detected by IRAS display flux ratio (f$_{60}$/f$_{100}$) values which are typical of SF galaxies. The two early spirals which display the larger flux ratio are part of S+S pairs with both members FIR emitting and having radial velocity separation between the members well below 200 km s$^{-1}$

Finally, on the basis of available data, we have analyzed the occurrence of AGN/SB phenomenon which amounts to 11% (computed on the total number of galaxies) and which we have shown is found mainly (88 %) in galaxy pairs having $\mid \Delta\ v_r \mid \leq$ 200 km s$^{-1}$ most (78 %) of which showing signs of interaction. Detailed homogeneous spectroscopic data (Marinoni et al. in preparation) coupled with surface photometry (Zitelli et al. in preparation) will allow to draw definitive conclusion on this topic.

*Acknowledgements.* This work was supported by MIUR. S.M. acknowledges a fellowship by INAF-OAB. This research has made use of the NASA/IPAC Extragalactic Database (NED) and of the Lyon-Meudon Extragalactic Database (LEDA).